\documentclass[letterpaper,10pt]{article} 

\usepackage{osameet3} 
\usepackage{subfigure}
\usepackage{graphicx,epstopdf}
\usepackage[utf8]{inputenc}
\usepackage{glossaries}
\usepackage[USenglish]{babel}
\usepackage{comment}
\usepackage{nicefrac}
\usepackage{amsmath,amssymb}
\usepackage[colorlinks=true,bookmarks=false,citecolor=blue,urlcolor=blue]{hyperref} 
\usepackage[capitalise]{cleveref} 

\newacronym{CUT}{CUT}{channel under test}

\newacronym{KK}{KK}{Kramers-Kronig}
\newacronym{CSPR}{CSPR}{carrier-to-signal power ratio}
\newacronym{KKRX}{KKRx}{Kramers-Kronig Receiver}
\newacronym{SSBI}{SSBI}{signal-signal beat interference}

\newacronym{GS}{GS}{geometric shaping}
\newacronym{PS}{PS}{probabilistic shaping}
\newacronym{DSP}{DSP}{digital signal processing}
\newacronym{MIMO}{MIMO}{multiple-input multiple-output}
\newacronym{TDE}{TDE}{time domain equalizer}
\newacronym{FDE}{FDE}{frequency domain equalizer}
\newacronym{LMS}{LMS}{least means square}
\newacronym{DDLMS}{DD-LMS}{decision directed least means square}
\newacronym{FFE}{FFE}{feed-forward equalizer}
\newacronym{FBE}{FBE}{feedback equalizer}
\newacronym{BPS}{BPS}{blind phase search}

\newacronym{SMF}{SMF}{single-mode fiber}
\newacronym[plural=SSMFs]{SSMF}{SSMF}{standard single-mode fiber}
\newacronym[plural=FMFs]{FMF}{FMF}{few-mode fiber}
\newacronym{FMF12}{FMF12}{12 mode FMF}
\newacronym{MMF}{MMF}{multi-mode fiber}
\newacronym{SI}{SI}{step index}
\newacronym{GI}{GI}{graded index}
\newacronym{DCF}{DCF}{dispersion compensated fiber}

\newacronym{SDM}{SDM}{space division multiplexing}
\newacronym{MDM}{MDM}{mode division multiplexed}
\newacronym{WDM}{WDM}{wavelength division multiplexing}
\newacronym{DWDM}{DWDM}{dense wavelength division multiplexing}
\newacronym{LP}{LP}{linear polarized}
\newacronym[plural=MMUXs,firstplural=mode multiplexers]{MMUX}{MMUX}{mode multiplexer}
\newacronym{PL}{PL}{photonic lantern}
\newacronym{3DWG}{3DWG}{3D-waveguide}
\newacronym{MDL}{MDL}{mode dependent loss}
\newacronym{MDG}{MDG}{mode dependent gain}
\newacronym{DGD}{DGD}{differential group delay}
\newacronym{DMGD}{DMGD}{differential mode group delay}
\newacronym{QSM}{QSM}{quasi-single-mode}
\newacronym{GIMMF}{GI-MMF}{graded-index multi-mode fiber}

\newacronym{SSB}{SSB}{single side band}
\newacronym{QPSK}{QPSK}{quadrature phase shift keying}
\newacronym{QAM}{QAM}{quadrature amplitude modulation}
\newacronym{RRC}{RRC}{root-raised-cosine}
\newacronym{4D-64PRS}{4D-64PRS}{
four-dimensional 64-ary polarization-ring-switching}
\newacronym{DP}{DP}{dual-polarization}
\newacronym[\glslongpluralkey=states-of-polarization]{SOP}{SOP}{state-of-polarization}
\newacronym{PM}{PM}{polarization-multiplexed}

\newacronym{ECL}{ECL}{external cavity laser}
\newacronym{CW}{CW}{continuous wave}
\newacronym[plural=DFBs]{DFB}{DFB}{distributed feedback laser}

\newacronym[plural=DACs]{DAC}{DAC}{digital-to-analog converter}
\newacronym{ADC}{ADC}{analog-to-digital converter}
\newacronym{PRBS}{PRBS}{pseudo-random bit sequence}
\newacronym{LO}{LO}{local oscillator}
\newacronym{EDFA}{EDFA}{erbium-doped fiber amplifier}
\newacronym{MZM}{MZM}{Mach-Zehnder modulator}
\newacronym{DP-MZM}{DP-MZM}{dual-polarization Mach-Zehnder modulator}
\newacronym{ChUT}{ChUT}{channel under test}
\newacronym{WSS}{WSS}{wavelength selective switch}
\newacronym[plural=VOAs]{VOA}{VOA}{variable optical attenuator}
\newacronym[plural=PDCRXs]{PDCRX}{PDCRX}{polarization diverse coherent receiver}
\newacronym{DSO}{DSO}{digital storage oscilloscope}
\newacronym{ASE}{ASE}{amplified spontaneous emission}
\newacronym{PBS}{PBS}{polarization beam splitter}
\newacronym{PD}{PD}{photodiode}
\newacronym{AOM}{AOM}{acousto-optical modulator}
\newacronym{BPD}{BPD}{balanced photo-diode}
\newacronym{OMFT}{OMFT}{optical-multi-format transmitter}
\newacronym{DPIQ}{DP-IQM}{dual-polarization IQ-modulator}
\newacronym{ABC}{ABC}{automatic bias controller}
\newacronym{OTF}{OTF}{optical tunable filter}
\newacronym{LSPS}{LSPS}{loop-synchronous polarization scrambler}
\newacronym{OSA}{OSA}{optical spectrum analyzer}

\newacronym{OSNR}{OSNR}{optical signal to noise ratio}
\newacronym{BER}{BER}{bit error rate}
\newacronym{IL}{IL}{insertion loss}

\newacronym{SDFEC}{SD-FEC}{soft-decision forward error correction}
\newacronym{HDFEC}{HD-FEC}{hard-decision forward error correction}
\newacronym{FEC}{FEC}{forward error correction}
\newacronym{LDPC}{LDPC}{low-density parity-check code}

\newacronym{AIR}{AIR}{achievable information rate}
\newacronym{AR}{AR}{achievable rates}
\newacronym{MI}{MI}{mutual information}
\newacronym{GMI}{GMI}{generalized mutual information}
\newacronym{BICM}{BICM}{bit-interleaved coded modulation}

\newacronym{OVNA}{OVNA}{optical vector network analyzer}
\newacronym{NIR}{NIR}{near infrared}
\newacronym{CD}{CD}{chromatic dispersion}
\newacronym{OTDR}{OTDR}{optical time domain reflectometry}
\newacronym{OFDR}{OFDR}{optical frequency domain reflectometry}
\newacronym{GPU}{GPU}{graphics processing unit}
\newacronym{SVD}{SVD}{singular value decomposition}
\newacronym{WGN}{WGN}{white Gaussian noise}
\newacronym{AWGN}{AWGN}{additive white Gaussian noise}
\newacronym{PDL}{PDL}{polarization dependent loss}
\newacronym{SPS}{sps}{samples-per-symbol}
\newacronym{SE}{SE}{spectral efficiency}
\newacronym{DD}{DD}{decision-directed}
\newacronym{MMSE}{MMSE}{minimum mean square error}
\newacronym{MSE}{MSE}{mean squared error}
\newacronym{SNR}{SNR}{signal to noise ratio}
\newacronym{ML}{ML}{machine learning}

\newacronym[plural=NNs]{NN}{NN}{neural-network} 
\newcommand{\LP}[1]{LP\textsubscript{#1}}
\begin{document}

\title{Neural-network-based MDG and Optical SNR Estimation in SDM Transmission}

\author{Ruby S. B. Ospina\textsuperscript{1, 2}, Menno van den Hout\textsuperscript{2}, Sjoerd van der Heide\textsuperscript{2},\\Chigo Okonkwo\textsuperscript{2} and Darli A. A. Mello\textsuperscript{1}}
\address{$^{1}$School of Electrical and Computer Engineering, University of Campinas, Campinas, 13083-852, BR\\$^{2}$High-Capacity Optical Transmission Laboratory, Eindhoven University of Technology, 5600 MB, NL}
\email{e-mail: ruby@decom.fee.unicamp.br}

\copyrightyear{2021}

\begin{abstract}
We propose a neural network model for MDG and optical SNR estimation in SDM transmission. We show that the proposed neural-network-based solution estimates MDG and SNR with high accuracy and low complexity from features extracted after DSP.
\end{abstract}

\section{Introduction}

\Gls{SDM} is currently regarded as the only solution to cope with the exponential growth of data traffic in optical transmission networks \cite{arik2014mimo, winzer2014optical}. In coupled \gls{SDM} transmission, mode coupling 
can be compensated by \gls{MIMO} equalizers at the receiver. In contrast, \gls{MDG} generated in inline amplifiers cannot be compensated by \gls{DSP}. The random power variations of the guided modes induced by MDG turn the channel capacity into a random variable, reducing the average capacity and generating outages \cite{ho2011mode, winzer2011mimo, Mello2014}. The combined effect of MDG and \gls{ASE} noise generated in amplifiers poses fundamental performance limitations to high-capacity \gls{SDM} systems deployed at long distances.

MDG estimation based on the transfer function of MIMO equalizers has been widely used in recent works \cite{van2017138,rademacher202010} to assess the link quality. However, we show in \cite{Ospina2020OFC} that, as adaptive \gls{MIMO} equalizers typically use the \gls{MMSE} criterion \cite{faruk2017digital}, the \gls{MDG} estimation accuracy is affected by the \gls{SNR}, mainly for high levels of MDG and low SNRs \footnote{We denote SNR as the ratio of total signal optical power and total optical noise power in the channel bandwidth, considering all supported spatial and polarization modes.}.
Based on a known \gls{SNR}, we also propose and validate a correction factor to partially compensate for the \gls{MDG} estimation errors \cite{MennoECOC2020MDL,Ospina2020JLT}. However, measuring the \gls{SNR} at the coherent receiver input may not be feasible in particular scenarios, limiting the scope of the proposed solution. Furthermore, although SNR estimation can be easily carried out in polarization-multiplexed optical systems from the signal after polarization demultiplexing, in mode-multiplexed systems this task is not straightforward, as the output signal-to-interference-plus-noise ratio (SINR) is MDG-dependent.  

Currently, \gls{ML} techniques are being considered for optical performance monitoring in both single-mode \cite{saif2020machine} and mode-multiplexed systems \cite{saif2020optical}. Nevertheless, to the best of our knowledge, the study of the joint MDG and SNR estimation in SDM systems based on \gls{ML} has not been yet reported. In this paper, we propose a \gls{NN} model to estimate \gls{MDG} and SNR in coupled SDM transmission systems. The model is validated experimentally in a 3-mode transmission system with polarization multiplexing over a 32.5 m \gls{FMF} link.

\section{Feature extraction for neural-network-based MDG and SNR estimation}
\label{sec:dsp estimation}
The MDG of a link can be computed from the eigenvalues, $\lambda_{i}^{2}$, of $\mathbf{H}\mathbf{H}^{H}$, where $\mathbf{H}$ is the channel transfer matrix, and $(.)^{H}$ is the Hermitian transpose operator \cite{winzer2011mimo,ho2011mode}. The standard deviation of the overall MDG, $\sigma_{\mathrm{mdg}} = \mathrm{std(log(}\lambda_{i}^{2}))$, is widely used to quantify the accumulated MDG at the end of the link. For an unknown $\mathbf{H}$, $\sigma_{\mathrm{mdg}}$ is conventionally computed from the eigenvalues, $\lambda^{2}_{i_{\mathrm{MMSE}}}$, of $\mathbf{W}^{-1}_{\mathrm{MMSE}}(\mathbf{W}^{-1}_{\mathrm{MMSE}})^H$, where the inverse transfer matrix of the MIMO MMSE equalizer, $\mathbf{W}^{-1}_{\mathrm{MMSE}}$, is used as an estimate of $\mathbf{H}$. The transfer matrices $\mathbf{W}_{\mathrm{MMSE}}$ and $\mathbf{H}$ are related by \cite{McKay:2010}
\begin{equation}
\mathbf{W}_{\mathrm{MMSE}} = \left( \frac{\mathbf{I}}{\mathrm{SNR}} + \mathbf{H}^{H}\mathbf{H} \right)^{-1}\mathbf{H}^{H}, 
\label{Eq: wmmse}    
\end{equation}
where the $\mathrm{SNR}$ is computed at the coherent receiver input. From (\ref{Eq: wmmse}), the estimated $\sigma_{\mathrm{mdg}}$ depends clearly on $\mathrm{SNR}$ \cite{Ospina2020OFC}.

Conventionally, in single-mode coherent optical systems, the optical SNR can be easily estimated after DSP from the electrical SNR. In SDM systems, however, the combined effect of MDG and ASE noise complicates this task. In coherent optical systems that use MMSE equalizers, the so-called electrical SNR is actually the signal-to-noise-plus interference ratio, $\mathrm{SINR}$. The SINR in data stream $i$ can be calculated as \cite{McKay:2010}
\begin{equation}
\mathrm{SINR}_{i} = \frac{1}{\left[ \left( \mathbf{I} + \mathrm{SNR'} \; \mathbf{H}^{H}\mathbf{H}  \right)^{-1} \right]_{i,\, i}} - 1,
\label{Eq:SNRSINRrelation}
\end{equation}
where $[\;]_{i, \, i}$ indicates the i-th element in the main diagonal. To account for the implementation penalty present in practical receivers, and mitigate imprecisions at high SNRs, the $\mathrm{SNR'}$ in (\ref{Eq:SNRSINRrelation}) is defined as \large{$ \nicefrac{1}{\left( \mathrm{SNR^{-1}} + \mathrm{SINR_{imp}^{-1}}   \right)} $}, \normalsize where $\mathrm{SINR_{imp}}$ is computed in a practical receiver in the absence of MDG and \gls{ASE} noise.

\begin{figure}[t]
	\centering
		\includegraphics[width=14.05cm]{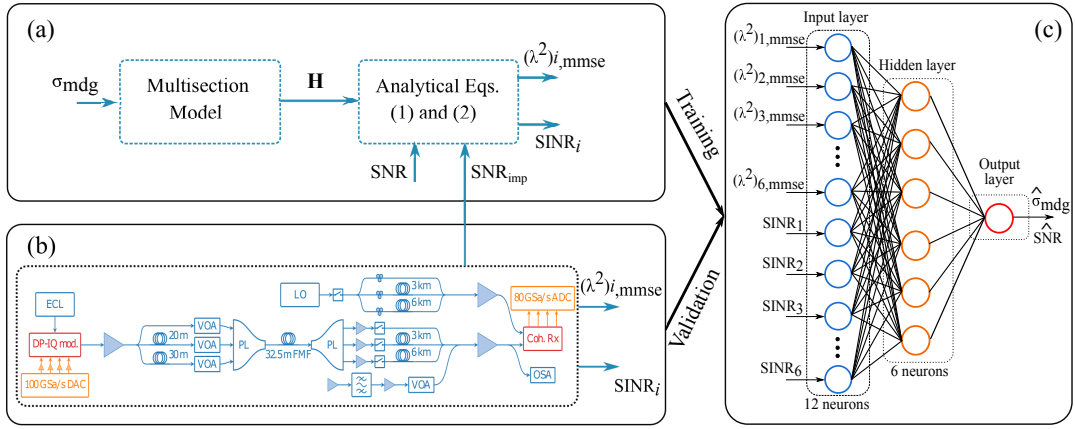}%
	\caption{\small{NN-based solution block diagram. (a) Analytic training set generation. (b) Experimental validation set generation. (c) Proposed NN. The algorithm applies two separate networks for $\mathrm{\sigma_{mdg}}$ and SNR estimation.}}
	\label{fig:generaldiagram}
\end{figure}

In this paper, we propose a NN  model to estimate $\mathrm{\sigma_{mdg}}$ and $\mathrm{SNR}$ from features extracted after DSP. The block diagram of the proposed solution is depicted in \cref{fig:generaldiagram}.
The training dataset is generated according to Inset (a). Using the multisection model presented in \cite{ho2011mode}, $6\times6$ matrices $\mathbf{H}$ are generated to simulate a 3-mode transmission with polarization multiplexing over a 2,500 km \gls{FMF} link with 0.2 dB $< \mathrm{\sigma_{mdg}}<$ 6.2 dB. For each $\mathbf{H}$, the $\mathrm{SNR}$ is swept from 5 dB to 22 dB to generate 6 $\lambda^{2}_{i_{MMSE}}$ values and 6 $\mathrm{SINR}_{i}$ values through equations (\ref{Eq: wmmse}) and (\ref{Eq:SNRSINRrelation}). The labelled set of $\lambda^{2}_{i_{MMSE}}$ and $\mathrm{SINR}_{i}$ values is fed into Inset (c) as input features for NN training. The \gls{NN}, implemented in \textit{keras/tensorflow}, receives 6 $\lambda^{2}_{i_{MMSE}}$ values and 6 $\mathrm{SINR}_{i}$ values, and provides an estimate of $\sigma_{\mathrm{mdg}}$ or $\mathrm{SNR}$. A hidden layer with 6 neurons, and an output layer with 1 neuron, learn the relation between the input features and the output based on the training samples generated analytically. The \gls{NN} is trained using Adam optimizer \cite{adam2014} during 500 epochs and a batch size of 5 samples.

 The NN model is validated using experimental data captured from the 3-mode transmission setup depicted in Inset (b). Three linearly polarized modes, \LP{01}, \LP{11a} and \LP{11b}, each one with two polarizations, are transmitted over 32.5 m of \gls{FMF}. \Glspl{VOA} are used to control the $\sigma_{\mathrm{mdg}}$ of the link. The optical \gls{SNR} is varied at the coherent receiver input by a noise loading stage. The $\mathrm{SNR}$ is computed as $\mathrm{ SNR = OSNR \, (T_{s} \times 12.5 \,GHz)}$ where $\mathrm{T_{s} = 40 \, ps}$ is the symbol time, and the OSNR is the traditional optical signal to noise ratio computed by an optical spectrum analyzer at the 12.5 GHz bandwidth. Additional details of the experimental setup can be found in \cite{Ospina2020JLT}. After DSP, the eigenvalues, $\lambda^{2}_{i_{MMSE}}$, are computed at each frequency of $\mathbf{W}_{\mathrm{MMSE}}$ and averaged across the signal band. The $\mathrm{SINR}_{i}$ is computed from each one of the 6 equalized data streams using a single-coefficient least-squares (LS) estimator \cite{X.Wautelet2027}.
 
\section{Neural-network-based $\mathrm{\sigma_{mdg}}$ and $\mathrm{SNR}$ estimation results}

\begin{figure}[t]
\centering
    	\includegraphics[width=14cm]{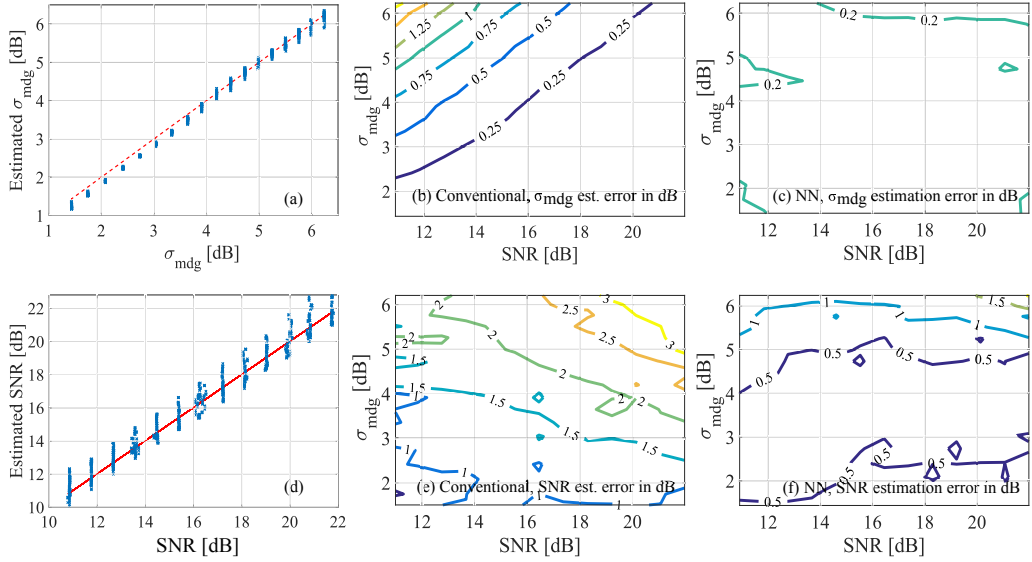}%
\caption{\small{(a) NN-estimated $\sigma_{\mathrm{mdg}}$ as a function of the actual $\sigma_{\mathrm{mdg}}$. (b) $\sigma_{\mathrm{mdg}}$ estimation error in dB generated by the conventional method as a function of the actual $\sigma_{\mathrm{mdg}}$ and $\mathrm{SNR}$. (c) $\sigma_{\mathrm{mdg}}$ estimation error in dB generated by the NN as a function of the actual $\sigma_{\mathrm{mdg}}$ and $\mathrm{SNR}$.
(d) NN-estimated $\mathrm{SNR}$ as a function of the actual $\mathrm{SNR}$. (e) $\mathrm{SNR}$ estimation error in dB generated by the conventional method as a function of the actual $\sigma_{\mathrm{mdg}}$ and $\mathrm{SNR}$. (f) $\mathrm{SNR}$ estimation error in dB generated by the NN as a function of the actual $\sigma_{\mathrm{mdg}}$ and $\mathrm{SNR}$.}}
\label{fig:nn_pred_actual}
\end{figure}

The NN is fed with 12,300 analytical samples. 90$\%$ of the samples are used for model training and the remainder for model testing. After training, model validation is performed from 936 experimental samples.
 Figs. \ref{fig:nn_pred_actual}a,d show the estimated  versus  actual values for $\sigma_{\mathrm{mdg}}$ and $\mathrm{SNR}$, respectively. The estimated values satisfactorily track the actual values, resulting in a \gls{MSE} of 0.019 for $\sigma_{\mathrm{mdg}}$ and 0.462 for $\mathrm{SNR}$. 

 Figs. \ref{fig:nn_pred_actual}b,e show the estimation error of the conventional method in dB, defined as the difference between the actual value and the estimated value. The conventional method for SNR estimation applies an LS estimator to the data flows after DSP, and for $\sigma_{\mathrm{mdg}}$ estimation uses the eigenvalues of the equalizer transfer function. The implementation penalty contribution is also taken into account to improve accuracy at high SNRs.   The conventional method provides a $\sigma_{\mathrm{mdg}}$ estimation error up to 1.75 dB at high MDG and low $\mathrm{SNR}$. In the case of $\mathrm{SNR}$, the estimation error achieves up to 3.3 dB at high levels of MDG and high $\mathrm{SNR}$.
 Figs. \ref{fig:nn_pred_actual}c,f show the estimation error in dB for $\sigma_{\mathrm{mdg}}$ and $\mathrm{SNR}$, respectively, for the NN solution.
Only a small residual $\sigma_{\mathrm{mdg}}$ estimation error of 0.2 dB is observed for the evaluated configurations. For $\mathrm{SNR}$, a residual estimation error up to 1.5 dB is observed at very high $\sigma_{\mathrm{mdg}}$ and only in a small region of $\mathrm{SNR}$. On most of the grid, the $\mathrm{SNR}$ estimation error is lower than 0.5 dB.

\section{Conclusion}

We propose a NN model to estimate MDG and SNR in SDM systems with coupled channels, based on features extracted after DSP. The proposed model is evaluated in an experimental 3-mode transmission setup with polarization multiplexing. The results show that the NN-based solution estimates both MDG and SNR with high accuracy and low complexity, largely exceeding the performance  provided by conventional methods.

\vspace{2mm}
\noindent
\emph{\footnotesize{This work was partially supported by FAPESP under grants 2018/25414-6, 2017/25537-8, 2015/24341-7, 2015/ 24517-8, by the TU/e-KPN Smart Two project and by the NWO Gravitation Program on Research Center for Integrated Nanophotonics (Grant Number 024.002.033).}}


\begin{thebibliography}{99}
\small

\bibitem{winzer2014optical} P. J. Winzer \textit{et al.}, ``Optical {MIMO-SDM} system capacities," Proc. of OFC, paper Th1J.1, (2014).

\bibitem{arik2014mimo} S. O. Arik \textit{et al.}, ``{MIMO} signal processing for mode-division...," IEEE Sig. Proc. Magazine, \textbf{31}, pp. 25-34, (2014).


\bibitem{ho2011mode} K.-P. Ho \textit{et al.}, ``Mode-dependent loss and gain: statistics and...," Optics express, \textbf{19}, pp. 16612-16635, (2011).

\bibitem{winzer2011mimo} P. J. Winzer \textit{et al.}, ``{MIMO} capacities and outage probabilities in ...," Optics express, \textbf{19.17}, pp. 16680-16696, (2011).

\bibitem{Mello2014} D. A. A. Mello, et al. “Impact of polarization and mode-dependent...,” J. Lightw. Technol., \textbf{38}, pp. 303–318, 2020.

\bibitem{van2017138} J. v. Weerdenburg \textit{et al.}, ``138 {Tbit/s} transmission over 650 km graded-index...," Proc. of ECOC, pp. 1-3, (2017).

\bibitem{rademacher202010} G. Rademacher \textit{et al.}, ``10.66 {P}eta-{B}it/s Transmission over a...," Proc. of OFC, paper Th3H.1, (2020). 

\bibitem{Ospina2020OFC} R. S. B. Ospina \textit{et al.}, ``DSP-based Mode-dependent Loss and Gain...," Proc. of OFC, paper W2A47.1, (2020). 

\bibitem{faruk2017digital} M.S. Faruk \textit{et al.}, ``Digital signal processing for coherent...," J. Lightw. Technol., \textbf{35}, pp. 1125-1141, (2017).

\bibitem{MennoECOC2020MDL} M. v.d. Hout \textit{et al.}, ``Experimental validation of MDL emulation...," Proc. of ECOC, (2020).

\bibitem{Ospina2020JLT} R. S. B. Ospina \textit{et al.}, ``Mode-dependent Loss and Gain Estimation...," J. Lightw. Technol., (2020). 


\bibitem{saif2020machine}W. S, Saif \textit{et al.}, ``Machine Learning Techniques for Optical Performance...," IEEE Comm. Surveys \& Tutorials, (2020).

\bibitem{saif2020optical} W. S, Saif \textit{et al.}, ``Optical Performance Monitoring in Mode Division Multiplexed...," J. Lightw. Technol., (2020).

\bibitem{McKay:2010}M. R. {McKay} \textit{et al.}, ``Achievable sum rate of MIMO MMSE...," IEEE T. Inform. Theory, \textbf{56}, pp. 396-410, (2010).

\bibitem{adam2014}D. P. Kingma, and J. Ba, ``Adam: A method for stochastic optimization." arXiv preprint arXiv:1412.6980 (2014).


\bibitem{X.Wautelet2027}X. Wautelet \textit{et al.}, ``Comparison of EM-based algorithms for MIMO..." IEEE T. on comm. 55.1 (2007): 216-226.

\end{thebibliography}
\end{document}